\documentclass[conference, a4paper]{IEEEtran}
\IEEEoverridecommandlockouts

\usepackage{cite}
\usepackage{amsmath,amssymb,amsfonts}
\usepackage{algorithmic}
\usepackage{graphicx}
\usepackage{textcomp}
\usepackage{xcolor}
\usepackage{booktabs}
\usepackage{multirow}
\usepackage{subfigure}

\def\BibTeX{{\rm B\kern-.05em{\sc i\kern-.025em b}\kern-.08em
    T\kern-.1667em\lower.7ex\hbox{E}\kern-.125emX}}
\begin{document}

\title{Explicit Residual-Based Scalable Image Coding \\ for Humans and Machines\\
}

\author{\IEEEauthorblockN{Yui Tatsumi}
\IEEEauthorblockA{\textit{Graduate School of FSE,} \\
\textit{Waseda University}\\
Tokyo, Japan \\
yui.t@fuji.waseda.jp}
\and
\IEEEauthorblockN{Ziyue Zeng}
\IEEEauthorblockA{\textit{Graduate School of FSE,} \\
\textit{Waseda University}\\
Tokyo, Japan \\
zengziyue@fuji.waseda.jp}
\and
\IEEEauthorblockN{Hiroshi Watanabe}
\IEEEauthorblockA{\textit{Graduate School of FSE,} \\
\textit{Waseda University}\\
Tokyo, Japan \\
hiroshi.watanabe@waseda.jp}
}

\maketitle

\begin{abstract}
Scalable image compression is a technique that progressively reconstructs multiple versions of an image for different requirements. In recent years, images have increasingly been consumed not only by humans but also by image recognition models. This shift has drawn growing attention to scalable image compression methods that serve both machine and human vision (ICMH). Many existing models employ neural network-based codecs, known as learned image compression, and have made significant strides in this field by carefully designing the loss functions. In some cases, however, models are overly reliant on their learning capacity, and their architectural design is not sufficiently considered. In this paper, we enhance the coding efficiency and interpretability of ICMH framework by integrating an explicit residual compression mechanism, which is commonly employed in resolution scalable coding methods such as JPEG2000. Specifically, we propose two complementary methods: Feature Residual-based Scalable Coding (FR-ICMH) and Pixel Residual-based Scalable Coding (PR-ICMH). These proposed methods are applicable to various machine vision tasks. Moreover, they provide flexibility to choose between encoder complexity and compression performance, making it adaptable to diverse application requirements. Experimental results demonstrate the effectiveness of our proposed methods, with PR-ICMH achieving up to 29.57\% BD-rate savings over the previous work.

\end{abstract}


\begin{figure}[t]
  \centering
\includegraphics[width=\hsize]
{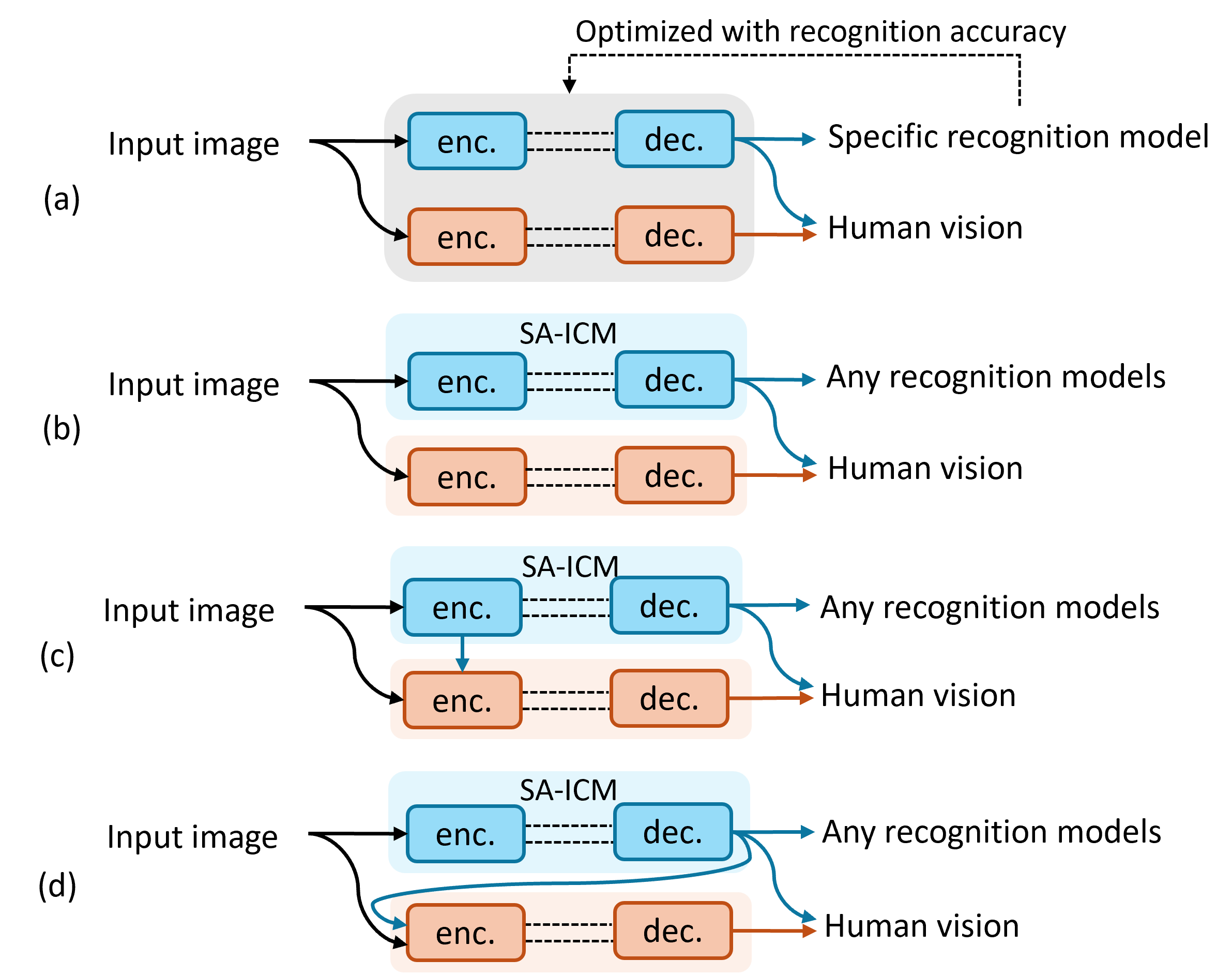}
  \caption{Overview of scalable image coding pipelines. (a) Conventional task-specific method, (b) ICMH-FF using decoder-side feature fusion, (c) FR-ICMH introducing encoder-side feature residual modeling, (d) PR-ICMH employing pixel-level residual compression.}
  \label{fig:codecs}
\end{figure}

\section{Introduction}
With the rapid advancement of deep learning, recognition models have become integral to a wide range of real-world applications, including traffic monitoring, agricultural management, camera surveillance, and industrial machine vision. In these scenarios, images are primarily analyzed by recognition models with occasional human inspection. To achieve scalable frameworks that integrate both human and machine vision, bridging the gap between these two domains has become a critical research challenge.


Various scalable approaches have been proposed to achieve image compression for machine and human vision (ICMH) \cite{ICMH-FF} - \cite{Shi}. While these methods provide efficient structures for scalability, most rely on Learned Image Compression (LIC) optimized for specific machine vision tasks. Hence, these scalable compression methods require independent training for each task, as shown in Fig.\ref{fig:codecs}(a). This task dependency may reduce flexibility in real-world applications where multiple recognition tasks are involved.

A notable attempt to address this challenge has been proposed in ICMH-FF \cite{ICMH-FF}, which introduces a task-agnostic scalable coding architecture. The overview of the framework is illustrated in Fig.\ref{fig:codecs}(b).
This method is based on the idea that the information in the image required for humans includes the information needed for machine vision. Specifically, it employs a machine-oriented LIC model that is independent of downstream tasks, alongside a separate LIC model to compress additional information required for human perception. On the decoder side, the features from both models are merged to reconstruct human-oriented images.
However, these models remain completely separated on the encoder side. This separation forces the encoder to infer the residual information to be compressed, resulting in a strong dependence on its learning capacity. Such an implicit mechanism leads to black-box behavior and may degrade compression efficiency as well as interpretability. Although designs that rely on implicit residual estimation are effective in reducing the encoder's computational burden, they stand in contrast to traditional and effective scalable compression approaches, which explicitly compute and encode residual information at each layer.

In this paper, we advance the ICMH method through the integration of explicitly modeled residual information.
In particular, we propose two complementary methods: Feature Residual-based Scalable Coding (FR-ICMH) and Pixel Residual-based Scalable Coding (PR-ICMH). The overview of the pipelines for these proposed methods are shown in Fig.\ref{fig:codecs}(c) and (d), respectively.
In FR-ICMH, the residual between machine-oriented and human-required features is computed and compressed. 
PR-ICMH extends this to the pixel domain by directly compressing the difference between the reconstructed machine-oriented image and the original one.
While FR-ICMH is suitable for scenarios that require a lightweight encoder design, PR-ICMH achieves improved rate-distortion performance by utilizing a more detailed residual representation. This trade-off between encoder complexity and reconstruction quality enables the proposed methods to flexibly adapt to a wide range of application constraints without affecting the recognition pipeline.

\section{Related Work}

\subsection{Learned Image Compression and Image Coding for Machines}\label{LIC}
LIC is an essential technology that underlies compression for human visual perception and machine vision \cite{balle} - \cite{LIC-last}. Most state-of-the-art LIC methods build on the frameworks of J. Ballé \textit{et al.}\cite{balle} and D. Minnen \textit{et al.} \cite{autoregressive}. Channel-wise autoregressive model (Ch-ARM) \cite{ch-arm} was then proposed to improve the decoding speed. This model slices the latent representation along the channel dimension and predicts entropy parameters using previously decoded slices. 
Building upon Ch-ARM, LIC-TCM\cite{LIC-TCM} further enhances the compression performance.

More recently, LIC tailored specifically for machine vision tasks, which is referred to as Image Coding for Machines (ICM), have also been explored \cite{ROI-first} - \cite{omni}.
These ICM models include ROI-based \cite{ROI-first} - \cite{ROI-last}, Task-Loss-based \cite{task-first} - \cite{task-last}, and Region-Learning-based \cite{SA-ICM} - \cite{Delta-ICM} approaches. 
The first two depend on task-specific prior analysis or loss functions, thus require separate training for each recognition task.
In contrast, Region-Learning-based methods aim to achieve task-agnostic compression by preserving recognition-relevant spatial regions. For instance, SA-ICM \cite{SA-ICM} leverages the Segment Anything model\cite{sa} to learn to retain only object boundaries while discarding textures during compression. This also provides privacy protection by removing most facial details. The model is trained using the following loss function:
\begin{equation}
\label{eqn:saicm-loss}
\mathcal{L} = \mathcal{R}(y) + \mathcal{R}(z) + \lambda \cdot mse(x \odot m, \hat{x} \odot m). 
\end{equation}
In (\ref{eqn:saicm-loss}), $y$ and $z$ denote the output of the encoder and hyperprior-encoder of the LIC model, respectively. $\mathcal{R}(y)$ and $\mathcal{R}(z)$ represent the estimated bitrates of $y$ and $z$. $x$ is the original image, and $\hat{x}$ is the reconstructed image. $mse$ denotes the mean squared error function. $\lambda$ is a weighting parameter that controls the trade-off between bitrate and distortion. $m$ is the object mask obtained by the Segment Anything model. The model architecture is based on LIC-TCM with Ch-ARM for entropy modeling.

\subsection{Scalable Image Coding for Humans and Machines}\label{ICMH}
Scalable coding for human and machine vision has emerged as an important direction in recent research.
H. Choi \textit{et al.}~\cite{Choi} proposed a dual-stream framework that splits the latent representation extracted from the input image into machine- and human-oriented components. These two streams are transmitted separately, and jointly decoded to reconstruct a human-oriented image.
More recently, Adapt-ICMH~\cite{ECCV2024} introduced a plug-and-play spatial-frequency modulation adapter for converting a human-oriented LIC into a machine-oriented codec. Scalability is achieved by operating either with the adapter for machine-oriented coding or without it for human-oriented coding. Notably, the adapter is compatible with any LIC architecture.
Although these methods are effective in terms of scalability and compression performance, they are trained to optimize recognition accuracy and therefore remain task-specific.

To address this limitation, ICMH-FF has been proposed. In this framework, SA-ICM is utilized for machine vision, while the additional LIC model provides complementary information required for human-oriented reconstruction. These two models are integrated through a feature fusion network implemented on the decoder side. Leveraging the channel-wise structure of Ch-ARM, this network performs channel-wise addition for overlapping slices and directly forwards the remaining slices from the SA-ICM stream. 
Through this architecture, ICMH-FF achieves scalability, adaptability to various recognition tasks, as well as effective compression performance.
However, as the connection between the two models is limited to the decoder side, the integration scheme leaves room for reconsideration.

\subsection{Compression Methods Using Residual Connection}
Residual connection-based coding has long been one of the core components of efficient image and video compression \cite{HEVC} - \cite{COMPASS}. In traditional standards such as HEVC\cite{HEVC} and VVC\cite{VVC}, the residual between the input frame and its predicted frame is encoded to achieve high compression efficiency. G. Lu \textit{et al.}~\cite{Lu} extended this principal to a learning-based video compression framework.

Residuals have also been leveraged within image compression methods aimed at resolution scalability. A typical approach first decodes a low-resolution image in the base layer, and subsequently reconstructs higher-resolution image by compressing the residual information in the enhancement layer.
A representative example is JPEG2000\cite{JPEG2000}, which utilizes wavelet-based decomposition to enable scalable encoding. 
Y. Mei  \textit{et al.}\cite{Y.Mei} proposed a learning-based scalable image compression model with one base layer and multiple enhancement layers. In this model, to reduce the inter-layer redundancy, the latent representations from the prior layers are utilized to predict the latents in the current enhancement layer, and then the prediction residuals are compressed. 
COMPASS\cite{COMPASS} enhances this approach with support for arbitrary scale factors. This method leverages an inter-layer arbitrary scale prediction module called LIFF and a residual compression module across all enhancement layers. They are applied repeatedly as the scale factors increase, ensuring efficient compression at various resolution levels.
These previous compression methods demonstrate that residual connections contribute significantly to achieving scalability in image compression.

\begin{figure}[t]
  \centering
\includegraphics[width=\hsize]
{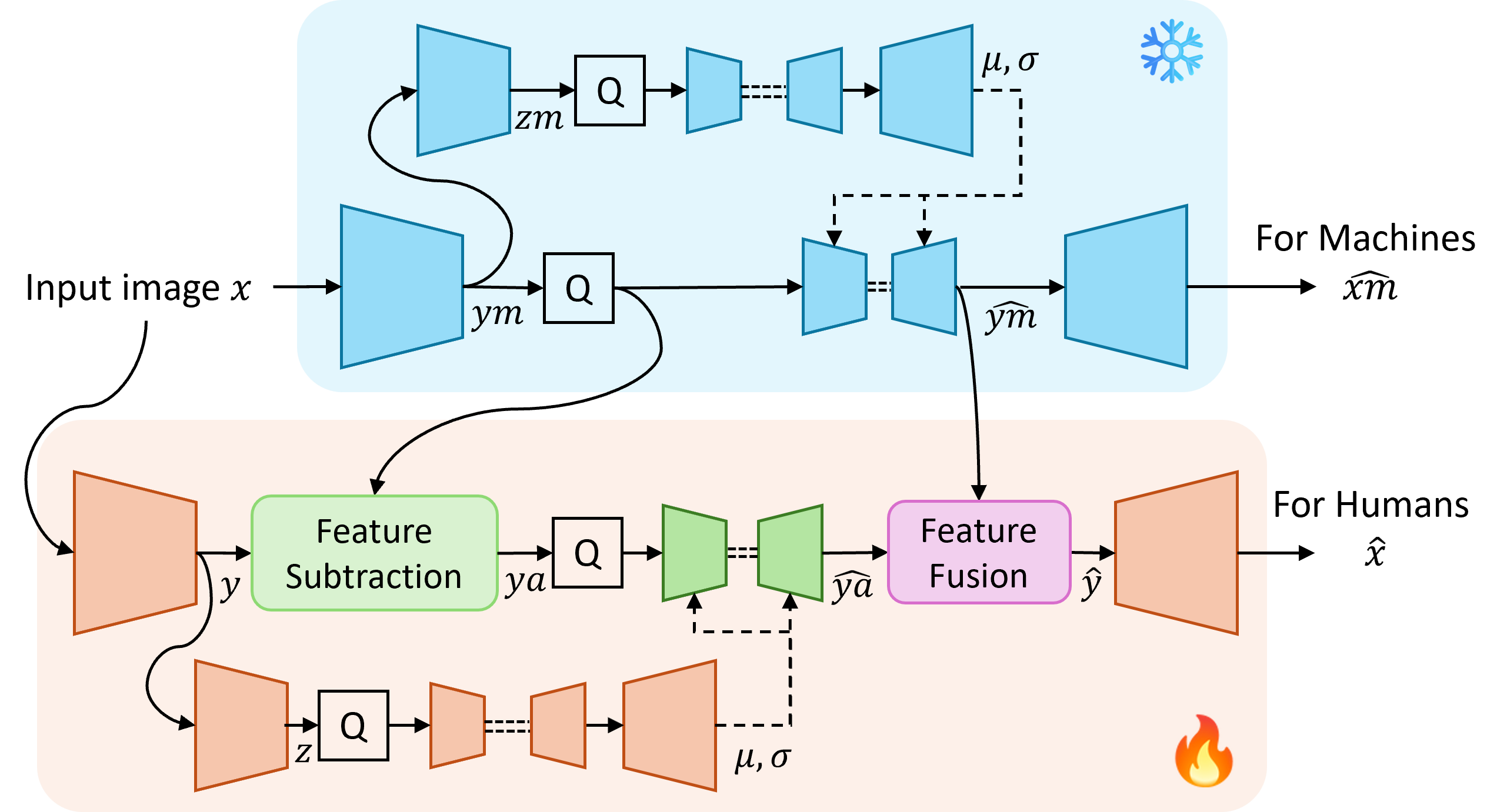}
  \caption{Overall architecture of the proposed FR-ICMH, which computes and encodes feature-level residuals between fixed SA-ICM (upper row) and an additional LIC (lower row). The residuals are subsequently fused with the quantized features from SA-ICM to reconstruct human-oriented images.}
  \label{fig:frc-model}
\end{figure}

\begin{figure}[t]
  \centering
\includegraphics[width=0.9\hsize]
{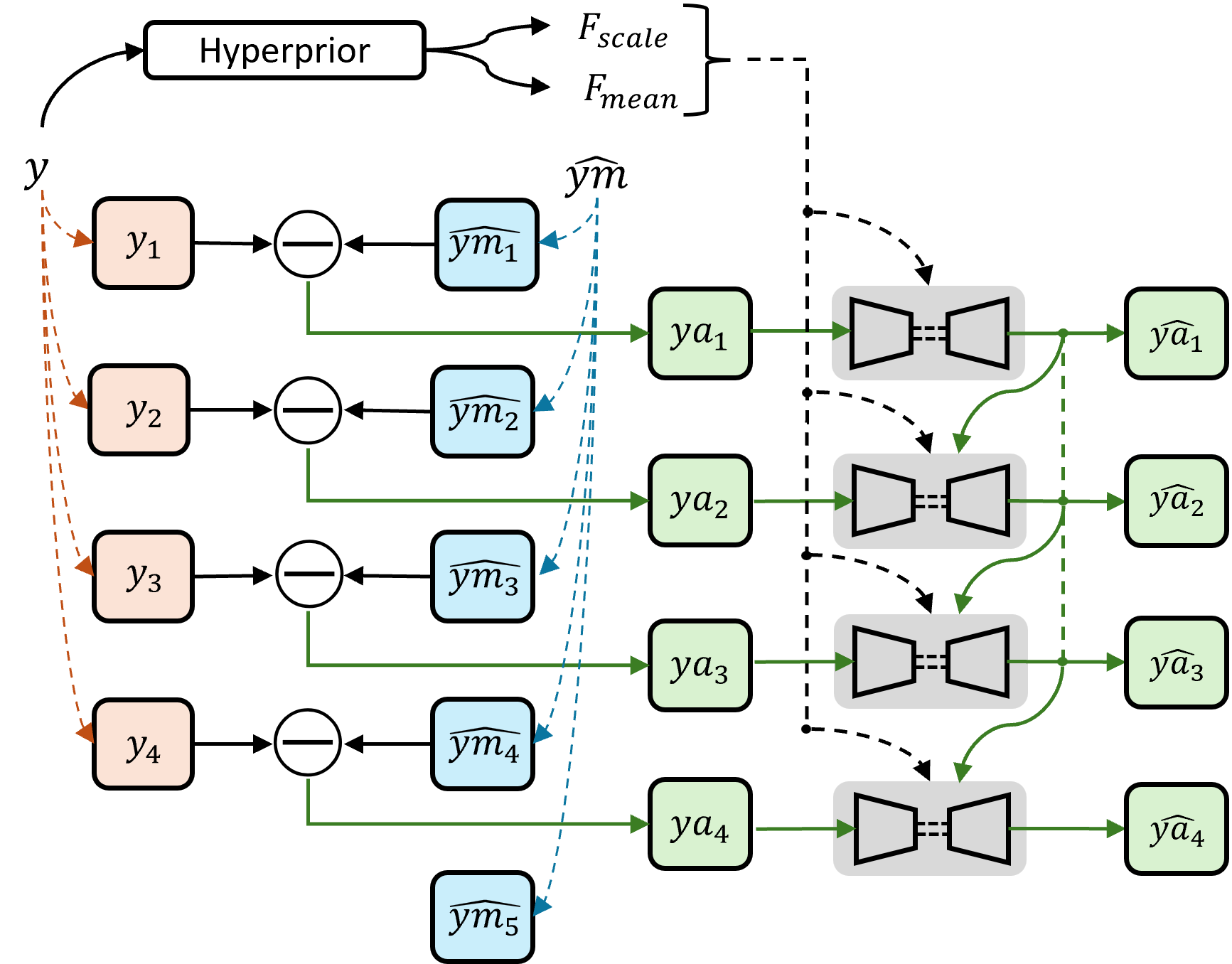}
  \caption{Structure of the feature subtraction network in FR-ICMH, which computes residuals between features from SA-ICM and LIC for each slice.}
  \label{fig:feature-subtraction}
\end{figure}
\section{Proposed Method}

\subsection{Overview}

\begin{figure}[t]
  \centering
\includegraphics[width=\hsize]
{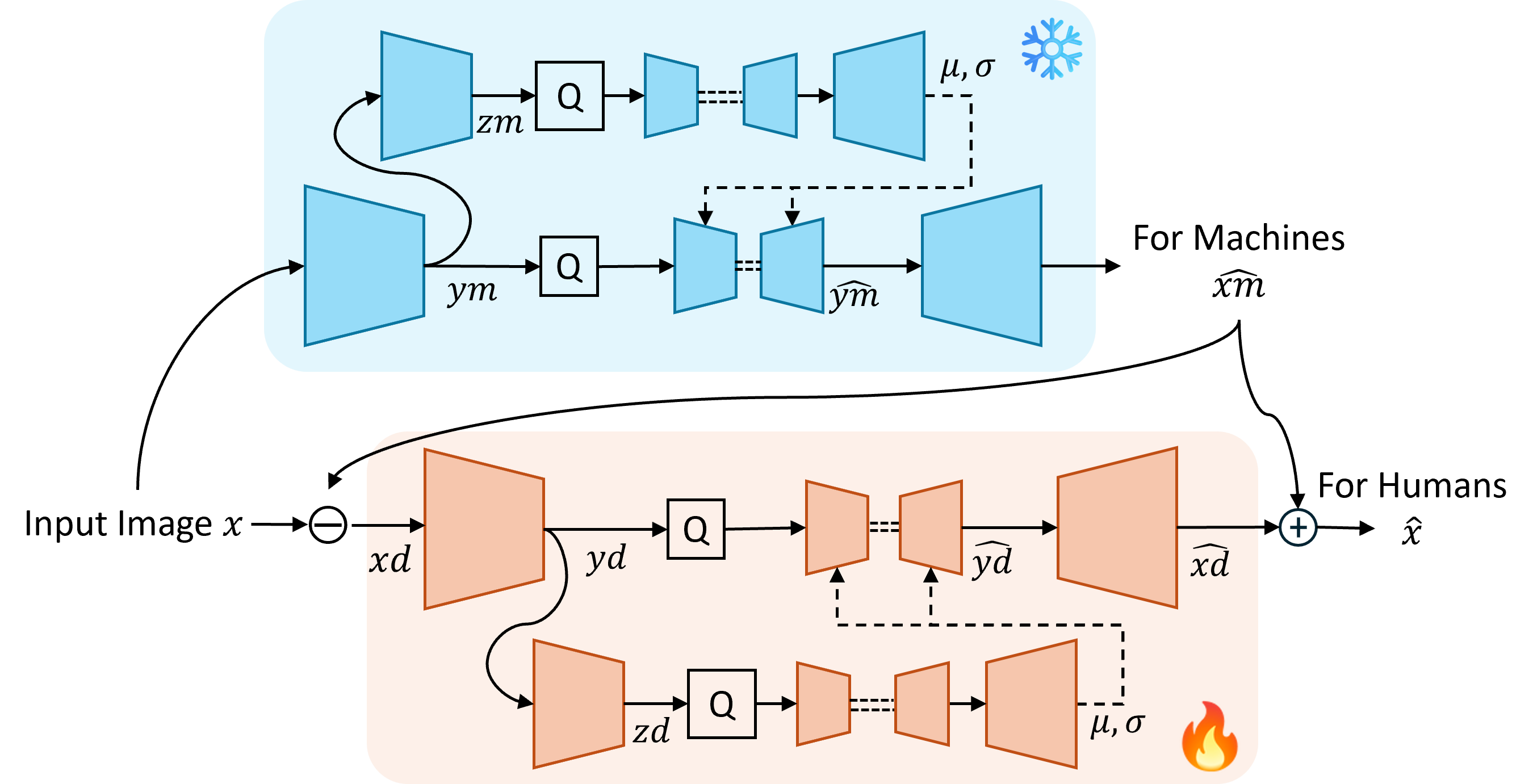}
  \caption{Overall architecture of the proposed PR-ICMH, which computes pixel-level residuals between the machine-oriented output of fixed SA-ICM (upper row) and the original image. The residuals are encoded by an additional LIC (lower row) and added back to reconstruct human-oriented images.}
  \label{fig:prc-model}
\end{figure}

\begin{figure*}[t]
  \centering
\includegraphics[width=\hsize]
{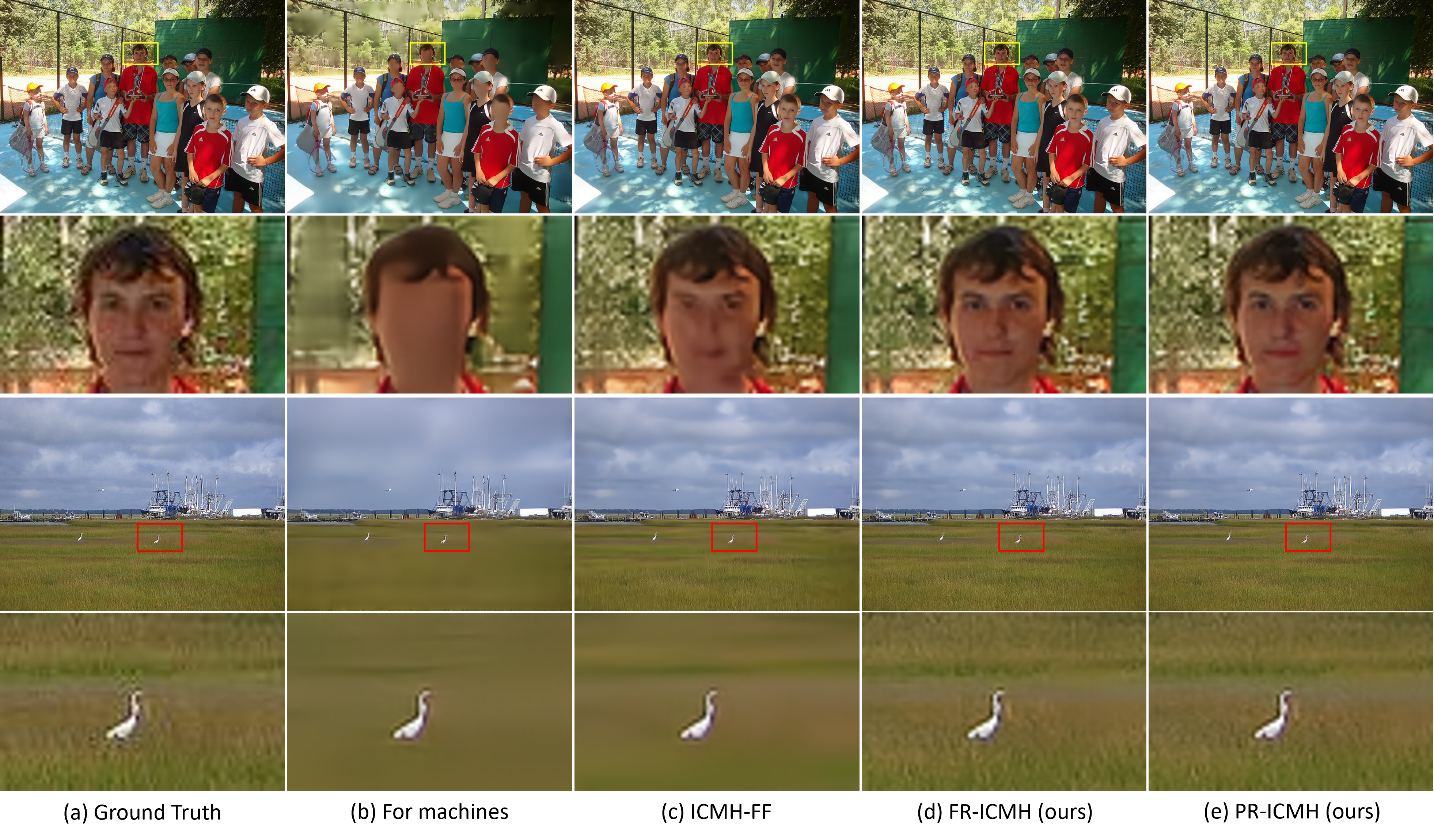}
  \caption{Examples of reconstructed images for machine and human vision. (a) Ground truth, (b) Machine-oriented reconstruction by SA-ICM ($N_m = 5$), (c) Human-oriented reconstruction by ICMH-FF ($N_a = 4$), (d) Human-oriented reconstruction by the proposed FR-ICMH ($N_a = 4$), and (e) Human-oriented reconstruction by the proposed PR-ICMH ($N_a = 4$). All reconstructions are obtained with $\lambda = 0.05$.}
  \label{fig:outputs}
\end{figure*}

We propose a residual-based scalable image coding framework that explicitly models the residual information required for human-oriented reconstruction. Our key idea is to avoid the encoder's implicit feature selection by directly providing the residual information as the compression target. By operating on clearly defined residuals, we aim to improve both compression efficiency and interpretability. We introduce two complementary methods:
\begin{itemize}
    \item \textbf{Feature Residual-based Scalable Coding (FR-ICMH)}: Computes and compresses the difference between human-oriented and machine-oriented features by utilizing a feature subtraction network.
    \item \textbf{Pixel Residual-based Scalable Coding (PR-ICMH)}: Directly compresses the pixel-level difference between the original image and the machine-oriented image.
\end{itemize}

Similar to ICMH-FF, the proposed methods are robust to variations in image recognition models, as they utilize a machine-oriented LIC model that remains independent of these models.
Furthermore, these methods offer distinct trade-offs: PR-ICMH prioritizes compression efficiency at the cost of higher encoder complexity, while FR-ICMH reduces complexity with a slight loss in performance. This enables the framework to adapt to various application scenarios.

\subsection{Feature Residual-based Scalable Coding}
The architecture of FR-ICMH is shown in Fig.\ref{fig:frc-model}. In this method, SA-ICM is utilized as the ICM model, while LIC-TCM is utilized as the additional LIC model for residual information. SA-ICM first encodes the input image into latent features $ym$, which are then quantized. Simultaneously, the additional LIC encoder processes the input image to produce $y$. The latent features from SA-ICM and the additional LIC model are divided along the channel dimension into $N_m$ and $N_a$ slices, respectively, denoted as $\{ym_1, ym_2, \ldots, ym_{N_m}\}$ and $\{y_1, y_2, \ldots, y_{N_a}\}$, where $1 \leq N_a \leq N_m$. By setting the number of slices $N_a$ smaller than $N_m$, the number of channels dedicated to residual information is decreased and computational cost can be reduced. 


To support residual computation between features with different numbers of slices, a feature subtraction network is employed to ensure compatibility. As shown in Fig.~\ref{fig:feature-subtraction}, in this network, the residual features $ya$ are computed slice-by-slice between corresponding slices of $y$ and quantized $ym$ as:
\begin{equation}
\label{eqn:feature-subtraction}
ya_k = y_k - \hat{ym_k}, \quad \text{where } k = 1,2,\ldots,N_a.
\end{equation}
In (\ref{eqn:feature-subtraction}), $ya_k$ represents a slice of residual features. These residual slices $ya_k$ are then compressed using Ch-ARM-based entropy modeling, maintaining the inherent decoding efficiency and scalability of the slice-wise structure. The decoder reconstructs the full latent representation by adding the decoded residual features $\hat{ya}$ and $\hat{ym}$ using the feature fusion network proposed in ICMH-FF. The feature fusion function is shown below:
\begin{gather}
\label{eqn:feature-fusion-1}
\hat{y_k} = 
\begin{cases}
\hat{ym_k} + \hat{ya_k} & (1 \leq k \leq N_a) \\
\hat{ym_k} & (N_a < k \leq N_m)
\end{cases} \\
\label{eqn:feature-fusion-2}
\hat{y} = conc(\hat{y_1}, \hat{y_2}, \ldots, \hat{y_{N_m}}).
\end{gather}
In (\ref{eqn:feature-fusion-1}) and (\ref{eqn:feature-fusion-2}), $\hat{y}$ denotes the input to the main decoder of additional LIC model. $conc$ stands for the concatenate function.
Only the additional LIC model for residual information is trained using the following loss function:
\begin{equation}
\label{eqn:frc-loss}
\mathcal{L} = \mathcal{R}(ya) + \mathcal{R}(z) + \lambda \cdot mse(x, \hat{x}). 
\end{equation}
In (\ref{eqn:frc-loss}), $ya$ and $z$ represent the outputs of the encoder and hyperprior-encoder of the additional LIC model.

\subsection{Pixel Residual-based Scalable Coding}
The architecture of PR-ICMH is illustrated in Fig.\ref{fig:prc-model}. Similar to FR-ICMH, SA-ICM is employed as the ICM model, and LIC-TCM serves as the additional LIC model for encoding the residual information. In this method, the pixel-level difference image $xd$ is computed by directly subtracting the machine-oriented reconstructed image $\hat{xm}$ from the original image $x$. This difference image is then compressed using an additional LIC model, which employs Ch-ARM for entropy modeling. Same as the architecture in FR-ICMH, the latent features are divided into $N_a$ slices, denoted as $\{yd_1, yd_2, \ldots, yd_{N_a}\}$. By setting $N_a$ to a smaller value, the number of channels used to represent $xd$ is reduced, thereby lowering the computational cost. At the decoder side, the final human-oriented reconstruction $\hat{x}$ is obtained by adding the machine-oriented image $\hat{xm}$ and the decoded difference image $\hat{xd}$. 
During the training, only the additional LIC model for difference image is trained with the following loss function:
\begin{equation}
\label{eqn:prc-loss}
\mathcal{L} = \mathcal{R}(yd) + \mathcal{R}(zd) + \lambda \cdot mse(xd, \hat{xd}). 
\end{equation}
In (\ref{eqn:prc-loss}), $yd$ and $zd$ denotes the outputs of encoder and hyperprior-encoder of LIC for difference image, respectively. 

\section{Experiment}

\subsection{Performance of Image Compression for Humans}

\begin{figure}[t]
  \centering
\includegraphics[width=\hsize]
{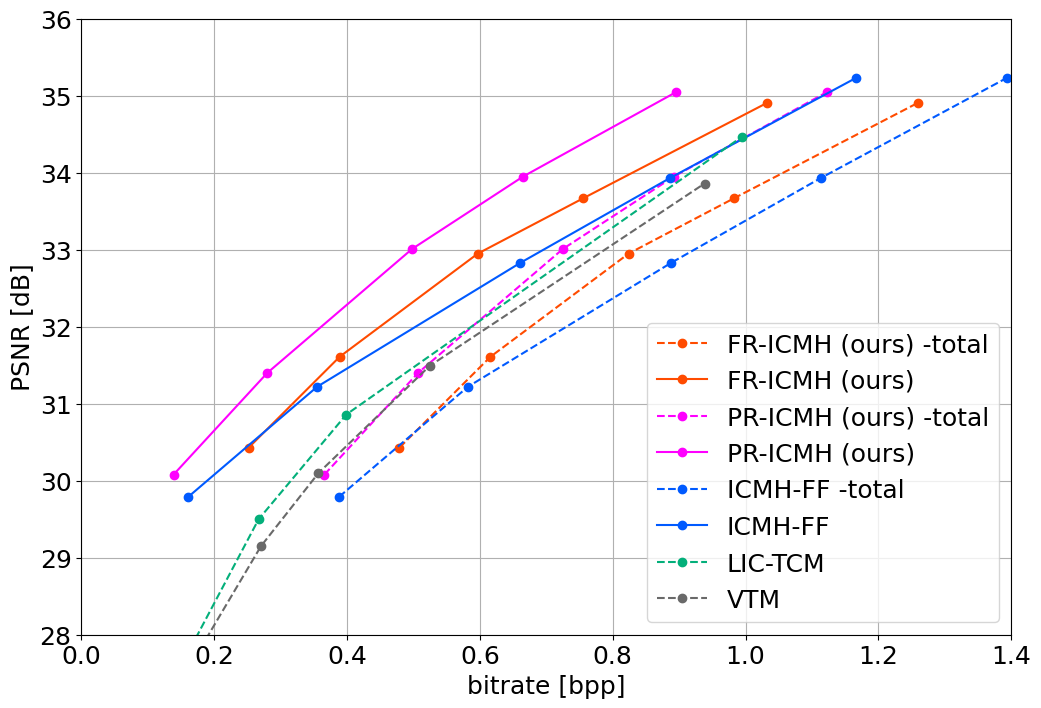}
  \caption{Rate-distortion curves for human-oriented image reconstruction with $N_a$ = 5.}
  \label{fig:b5}
\end{figure}

\begin{figure}
\centering
\subfigure[FR-ICMH]{%
\includegraphics[clip, width=0.48\hsize]{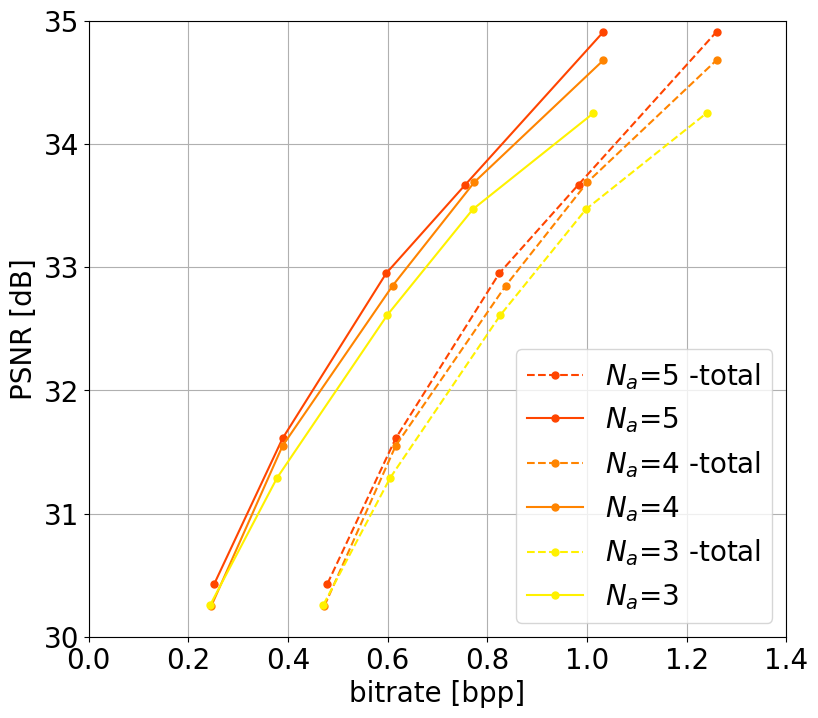}}%
\subfigure[PR-ICMH]{%
\includegraphics[clip, width=0.48\hsize]{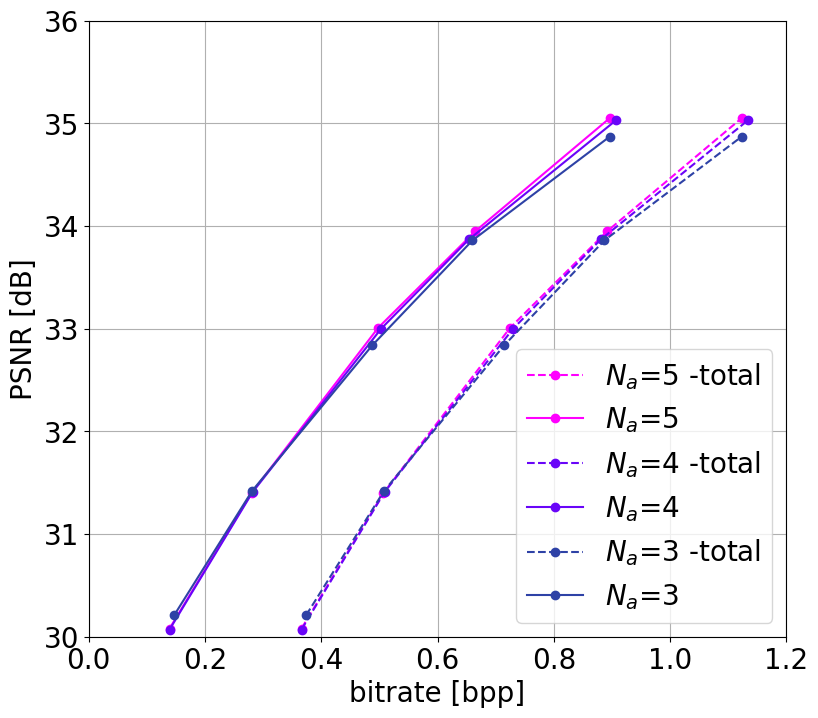}}%
\caption{Rate-distortion performance of (a) FR-ICMH and (b) PR-ICMH with different numbers of residual slices ($N_a = 3, 4, 5$).}
\label{fig:b-ablation}
\end{figure}

We evaluate the compression performance for humans of our proposed methods, FR-ICMH and PR-ICMH. For ICM, we utilize fixed SA-ICM model pre-trained with $\lambda=0.05$ according to the loss function defined in (\ref{eqn:saicm-loss}). For the additional LIC model for residual information, LIC-TCM is utilized. In both methods, only the additional LIC is trained. Since the parameters of SA-ICM are fixed, FR-ICMH and PR-ICMH achieve exactly the same performance as SA-ICM for all recognition tasks, including object detection and segmentation.
We train both FR-ICMH and PR-ICMH using their respective loss functions, (\ref{eqn:frc-loss}) and (\ref{eqn:prc-loss}). During the training, five different $\lambda$ values $\{0.005, 0.01, 0.02, 0.03, 0.05\}$ are utilized. COCO-train dataset is utilized for training and COCO-val is utilized for evaluation \cite{coco}. The number of feature slices for SA-ICM, $N_m$, is fixed to 5, while the additional LIC is trained with $N_a = \{3, 4, 5\}$ to evaluate the effect of number of slices of residual features.
Fig.~\ref{fig:outputs} shows that the proposed FR-ICMH and PR-ICMH provide clearer reconstructions of critical visual details, such as facial components and textures in grass, which are less distinct in ICMH-FF.

The rate-distortion curves for human-oriented images with $N_a = 5$ are presented in Fig.~\ref{fig:b5}. The solid lines indicate the bitrate of additional information only, while dashed lines represent the total bitrate (machine-oriented + additional). We compare our proposed methods with three existing methods: ICMH-FF, LIC-TCM, and VVC (VTM), a traditional compression codec.
FR-ICMH outperforms ICMH-FF especially at high bitrate regions, while PR-ICMH consistently outperforms across all bitrate ranges. Notably, PR-ICMH slightly outperforms LIC-TCM at high bitrate levels, despite its scalable structure. The performance degradation of FR-ICMH at low bitrates is primarily due to subtracting the high-quality machine-oriented features of SA-ICM ($\lambda=0.05$) from the human-oriented features. This subtraction yields a very small residual, and results in limited information recovery.

The BD-rate of our proposed methods compared to ICMH-FF for different number of slices, $N_a$, is shown in Table~\ref{tab:bd-rate}. For each value of $N_a$, the corresponding ICMH-FF model is utilized as the baseline. As shown in the table, both proposed methods outperform ICMH-FF across all configurations. In particular, PR-ICMH demonstrates consistently large gains regardless of the number of slices, achieving the best performance with a BD-rate reduction of 29.57\% when $N_a = 3$.

\subsection{Effect of Reduction in Number of Slices}
We further evaluate our proposed methods by investigating the effect of parameter reduction in the additional LIC model for residual information. Increasing the number of slices $N_a$ leads to a larger number of intermediate features, which in turn increases the number of model parameters. Fig.~\ref{fig:b-ablation} illustrates how the number of residual feature slices $N_a$ affects the rate-distortion performance of FR-ICMH and PR-ICMH.
While both methods show improved performance with higher $N_a$, the gain is marginal, especially for PR-ICMH. Table~\ref{tab:params} shows the corresponding increase in model parameters. These results indicate that a smaller model can be achieved by reducing the number of slices without significantly sacrificing compression performance.
\begin{table}[t]
  \centering
  \caption{BD-rate comparison with ICMH-FF for different values of the residual feature slices $N_a$}
  \begin{tabular}{cccc}
    \hline
    \multirow{2}{*}{Method} & \multicolumn{3}{c}{BD-rate(\%)} \\
    \cmidrule(lr){2-4}
                            & \(N_a = 3\) & \(N_a = 4\) & \(N_a = 5\) \\
    \hline \hline
    ICMH-FF                 & 0.00        & 0.00        & 0.00 \\ 
    FR-ICMH (ours)              & -2.43          & -4.23       & -7.93 \\
    PR-ICMH (ours)              & -29.57          & -26.06      & -26.78 \\
    \hline
  \end{tabular}
  \label{tab:bd-rate}
\end{table}

\begin{table}[t]
    \begin{center}
    \caption{Comparison between the size of residual information compression model and the number of slices $N_a$}
    \label{tab:params}
    \begin{tabular}{c|ccc}
        \hline
        $N_a$ & 3 & 4 & 5 \\ \hline 
        \hline
        Number of channels for residual information & 192 & 256 & 350 \\
        Number of parameters (M) & 58.7 & 67.0 & 76.6 \\
        \hline
    \end{tabular}
    \end{center}
\end{table}
\section{Conclusion}
In this paper, we propose a residual-based scalable image coding framework for human visual perception and machine vision. We introduce two complementary ICMH methods, Feature Residual-based (FR-ICMH) and Pixel Residual-based (PR-ICMH). By integrating explicitly modeled residual information, the proposed methods enhance the interpretability of the ICMH approach. Moreover, these methods are versatile across a wide range of application scenarios. This flexibility stems from their robustness to changes in downstream recognition models and their ability to balance computational cost and compression performance. Experimental results show that our proposed methods outperform the previous scalable method, ICMH-FF, with PR-ICMH achieving up to 29.57\% BD-rate reduction.
In future work, we aim to extend our residual-based framework to video coding and incorporate variable bitrate control for greater deployment adaptability.

\section*{Acknowledgment}
The results of this research were obtained from the commissioned research (JPJ012368C05101) by National Institute of Information and Communications Technology (NICT), Japan.


\begin{thebibliography}{00}
\bibitem{ICMH-FF} T. Shindo, T. Watanabe, Y. Tatsumi and H. Watanabe, ``Scalable Image Coding for Humans and Machines Using Feature Fusion Network,'' 2024 IEEE 26th International Workshop on Multimedia Signal Processing (MMSP), 2024, pp. 1-6.
\bibitem{Choi} H. Choi and I. V. Bajić, ``Scalable Image Coding for Humans and Machines,'' in IEEE Transactions on Image Processing, vol. 31, pp. 2739-2754, 2022.
\bibitem{ECCV2024} H. Li \textit{et al.}, ``Image Compression for Machine and Human Vision with Spatial-Frequency Adaptation,'' Computer Vision – ECCV 2024. ECCV 2024. Lecture Notes in Computer Science, vol. 15109, 2024, pp. 382-399.
\bibitem{}S. Sun, T. He and Z. Chen, ``Semantic Structured Image Coding Framework for Multiple Intelligent Applications,'' in IEEE Transactions on Circuits and Systems for Video Technology, vol. 31, no. 9, pp. 3631-3642, 2021.
\bibitem{sssic} N. Yan \textit{et al.}, ``SSSIC: Semantics-to-Signal Scalable Image Coding With Learned Structural Representations,'' in IEEE Transactions on Image Processing, vol. 30, pp. 8939-8954, 2021.
\bibitem{} H. Choi and I. V. Bajić, ``Scalable Video Coding for Humans and Machines,'' 2022 IEEE 24th International Workshop on Multimedia Signal Processing (MMSP), Shanghai, China, 2022, pp. 1-6.
\bibitem{} N. Le \textit{et al.}, ``Bridging the Gap Between Image Coding for Machines and Humans,'' 2022 IEEE International Conference on Image Processing (ICIP), 2022, pp. 3411-3415.
\bibitem{} S. Wang \textit{et al.}, ``Towards Analysis-Friendly Face Representation With Scalable Feature and Texture Compression,'' in IEEE Transactions on Multimedia, vol. 24, pp. 3169-3181, 2022.
\bibitem{} H. Hadizadeh and I. V. Bajic, ``Learned scalable video coding for humans and machines,'' arXiv preprint arXiv:2307.08978, 2023.
\bibitem{VVC+M} A. Harell, Y. Foroutan and I. V. Bajić, ``VVC+M : Plug and Play Scalable Image Coding for Humans and Machines,'' 2023 IEEE International Conference on Multimedia and Expo Workshops (ICMEW), 2023, pp. 200-205.
\bibitem{Andrade} A. de Andrade, A. Harell, Y. Foroutan and I. V. Bajić, ``Conditional and Residual Methods in Scalable Coding for Humans and Machines,'' 2023 IEEE International Conference on Multimedia and Expo Workshops (ICMEW), 2023, pp. 194-199.
\bibitem{} R. Feng, Y. Gao, X. Jin, R. Feng and Z. Chen, ``Semantically Structured Image Compression via Irregular Group-Based Decoupling,'' 2023 IEEE/CVF International Conference on Computer Vision (ICCV), 2023, pp. 17191-17201.
\bibitem{} Y. -H. Chen \textit{et al.}, ``TransTIC: Transferring Transformer-based Image Compression from Human Perception to Machine Perception,'' 2023 IEEE/CVF International Conference on Computer Vision (ICCV), 2023, pp. 23240-23250. 
\bibitem{} L. Liu, Z. Hu, Z. Chen and D. Xu, ``ICMH-Net: Neural Image Compression Towards both Machine Vision and Human Vision,'' in Proceedings of the 31st ACM International Conference on Multimedia, 2023, pp. 8047-8056. 
\bibitem{} Y. Wu, P. An, C. Yang and X. Huang, ``Scalable image coding with enhancement features for human and machine,'' Multimedia Systems, vol. 30, no. 77, 2024.
\bibitem{} T. Shindo, Y. Tatsumi, T. Watanabe and H. Watanabe, "Refining Coded Image in Human Vision Layer Using CNN-Based Post-Processing," 2024 IEEE 13th Global Conference on Consumer Electronics (GCCE), 2024, pp. 166-167.
\bibitem{} J. Wei \textit{et al.}, ``Layered and scalable image coding with semantic features for human and machine,'' Engineering Applications of Artificial Intelligence, vol. 155, 2025.
\bibitem{Shi} W. Shi, W. Yin, F. Tao and Y. Wen, ``Semantic Prior-Guided Scalable Image Coding,'' 2025 IEEE International Conference on Acoustics, Speech and Signal Processing (ICASSP), 2025, pp. 1-5.


\bibitem{balle} J. Ballé, D. Minnen, S. Singh, S. J. Hwang, and N. Johnston, ``Variational Image Compression with a Scale Hyperprior,'' International Conference on Learning Representations (ICLR), 2018, pp. 1-10.
\bibitem{autoregressive} D. Minnen, J. Ballé, and G. Toderici, ``Joint Autoregressive and Hierarchical Priors for Learned Image Compression,'' 32nd Conference on Neural Information Processing Systems (NeurIPS), 2018, pp.10794-10803.
\bibitem{ch-arm} D. Minnen and S. Singh, ``Channel-Wise Autoregressive Entropy Models for Learned Image Compression,'' 2020 IEEE International Conference on Image Processing (ICIP), 2020, pp. 3339-3343.
\bibitem{LIC-TCM} J. Liu, H. Sun and J. Katto, ``Learned Image Compression with Mixed Transformer-CNN Architectures,'' 2023 IEEE/CVF Conference on Computer Vision and Pattern Recognition (CVPR), 2023, pp. 14388-14397.
\bibitem{} J. Balle, V. Laparra, and E. P. Simoncelli, ``End-to-end optimized image compression,'' arXiv preprint arXiv:1611.01704, 2016.
\bibitem{LIC-last} Z. Cheng, H. Sun, M. Takeuchi and J. Katto, ``Learned Image Compression With Discretized Gaussian Mixture Likelihoods and Attention Modules,'' 2020 IEEE/CVF Conference on Computer Vision and Pattern Recognition (CVPR), 2020, pp. 7936-7945.

\bibitem{ROI-first} H. Choi and I. V. Bajic, ``High Efficiency Compression for Object Detection,'' 2018 IEEE International Conference on Acoustics, Speech and Signal Processing (ICASSP), 2018, pp. 1792-1796.
\bibitem{} Z. Huang, C. Jia, S. Wang and S. Ma, ``Visual Analysis Motivated Rate-Distortion Model for Image Coding,'' 2021 IEEE International Conference on Multimedia and Expo (ICME), 2021, pp. 1-6.
\bibitem{ROI-last} J. I. Ahonen, N. Le, H. Zhang, F. Cricri and E. Rahtu, ``Region of Interest Enabled Learned Image Coding for Machines,'' 2023 IEEE 25th International Workshop on Multimedia Signal Processing (MMSP), 2023, pp. 1-6.

\bibitem{task-first} N. Le, H. Zhang, F. Cricri, R. Ghaznavi-Youvalari and E. Rahtu, ``Image Coding For Machines: an End-To-End Learned Approach,'' 2021 IEEE International Conference on Acoustics, Speech and Signal Processing (ICASSP), 2021, pp. 1590-1594.
\bibitem{} N. Le, H. Zhang, F. Cricri, R. Ghaznavi-Youvalari, H. R. Tavakoli and E. Rahtu, ``Learned Image Coding for Machines: A Content-Adaptive Approach,'' 2021 IEEE International Conference on Multimedia and Expo (ICME), 2021, pp. 1-6.
\bibitem{task-last} X. Shen, H. Ou and W. Yang, ``Image Coding For Machine Via Analytics-Driven Appearance Redundancy Reduction,'' 2024 IEEE International Conference on Image Processing (ICIP), 2024, pp. 1883-1889.

\bibitem{SA-ICM} T. Shindo, K. Yamada, T. Watanabe and H. Watanabe, ``Image Coding For Machines With Edge Information Learning Using Segment Anything,'' 2024 IEEE International Conference on Image Processing (ICIP), 2024, pp. 3702-3708.
\bibitem{ccnc} T. Shindo, T. Watanabe, K. Yamada and H. Watanabe, ``Image Coding for Machines with Object Region Learning,'' 2024 IEEE 21st Consumer Communications \& Networking Conference (CCNC), 2024, pp. 1040-1041.
\bibitem{Delta-ICM} T. Shindo, T. Watanabe, Y. Tatsumi and H. Watanabe, "Delta-ICM: Entropy Modeling with Delta Function for Learned Image Compression," 2025 IEEE International Conference on Consumer Electronics (ICCE), 2025, pp. 1-6.

\bibitem{omni} R. Feng \textit{et al.}, ``Image Coding for Machines with Omnipotent Feature Learning,'' Computer Vision - ECCV 2022. ECCV 2022. Lecture Notes in Computer Science, vol. 13697, 2022, pp 510-528.


\bibitem{sa} A. Kirillov \textit{et al.}, ``Segment Anything,'' 2023 IEEE/CVF International Conference on Computer Vision (ICCV), 2023, pp. 3992-4003.

\bibitem{HEVC} High Efficiency Video Coding, Standard ISO/IEC 23008-2, ISO/IEC JTC 1, 2013.
\bibitem{VVC} Versatile Video Coding, Standard ISO/IEC 23090-3, ISO/IEC JTC 1, 2020.
\bibitem{Lu} G. Lu \textit{et al.}, ``DVC: An End-To-End Deep Video Compression Framework,'' 2019 IEEE/CVF Conference on Computer Vision and Pattern Recognition (CVPR), 2019, pp. 10998-11007.
\bibitem{JPEG2000} Information Technology—JPEG 2000—Image Coding System—Part1: Core Coding System, ISO/IEC 15 444-1, 2000.
\bibitem{Y.Mei} Y. Mei, L. Li, Z. Li and F. Li, ``Learning-Based Scalable Image Compression With Latent-Feature Reuse and Prediction,'' in IEEE Transactions on Multimedia, vol. 24, pp. 4143-4157, 2022.
\bibitem{COMPASS} J. Park, J. Lee and M. Kim, ``COMPASS: High-Efficiency Deep Image Compression with Arbitrary-scale Spatial Scalability,'' 2023 IEEE/CVF International Conference on Computer Vision (ICCV), 2023, pp. 12780-12789.

\bibitem{coco} T. Y. Lin \textit{et al.}, ``Microsoft COCO: Common Objects in Context,'' Computer Vision – ECCV 2014. ECCV 2014. Lecture Notes in Computer Science, vol. 8693, pp.740-755, 2014.

\end{thebibliography}
\end{document}